\documentclass{article}

\usepackage[nonatbib,final]{nips_2017}

\usepackage[usenames,dvipsnames]{xcolor}
\usepackage{amssymb}
\usepackage{amsmath}
\usepackage{amsthm}
\usepackage[utf8]{inputenc} 
\usepackage[T1]{fontenc}    
\usepackage{hyperref}         
\usepackage{url}                  
\usepackage{booktabs}       
\usepackage{amsfonts}       
\usepackage{nicefrac}         
\usepackage{microtype}      
\usepackage{hyperref}
\usepackage[numbers,square,sort&compress]{natbib}
\usepackage{graphicx}

\title{Vector-Valued Spectral Analysis of Space-Time Data}

\author{
  Dimitrios Giannakis \\
  Courant Institute \\
  of Mathematical Sciences,\\
  New York University,\\
  New York, NY 10012 \\
  \texttt{dimitris@cims.nyu.edu} \\
  \And
    Joanna Slawinska \\
    \textbf{Abbas Ourmazd}  \\
  Department of Physics, \\
  University of \\Wisconsin-Milwaukee,\\
  Milwaukee, WI 53211 \\
   \And
  Zhizhen Zhao  \\
  Department of Electrical \\ and Computer Engineering, \\
  University of Illinois \\ at Urbana-Champaign, \\
  Urbana, IL 61801 
}

\begin{document}

\maketitle

\begin{abstract}
Identifying coherent spatiotemporal patterns generated by complex dynamical systems is a central problem in many science and engineering disciplines. Here, we combine ideas from the theory of operator-valued kernels  with delay-embedding techniques in dynamical systems to develop of a method  for objective identification of spatiotemporal coherent patterns, without prior knowledge of the state space and/or the dynamical laws of the system generating the data. A key aspect of this method is that it operates on a space of vector-valued observables using a kernel measure of similarity that takes into account both temporal and spatial degrees of freedom (in contrast, classical kernel techniques such as PCA utilize aggregate measures of similarity between ``snapshots''). As a result, spectral decomposition of data via our approach yields a significantly more efficient and physically meaningful representation of complex spatiotemporal signals than conventional methods based on scalar-valued kernels. We demonstrate this behavior with applications to an oscillator model and sea surface temperature data from a climate model. 
\end{abstract}

\section{Introduction}

Fleeting spatiotemporal patterns are ubiquitous in nature, ranging from ultrafast molecular processes to slow geophysical convective flows, and ultraslow astrophysical phenomena  \cite{CrossHohenberg93,AhlersEtAl09,SlawinskaEtAl14,FungEtAl16}.  The objective identification and predictive modeling of such patterns remain longstanding challenges of wide significance.  The combination of modern mathematical analysis techniques with the accelerating capability to collect data presents a significant opportunity to improve our capability to understand, predict, and ultimately control pattern-forming systems.

Many of the data decomposition techniques developed over the years \cite{AubryEtAl91,ScholkopfEtAl98,BelkinNiyogi03,CoifmanEtAl05,HeinEtAl05,CoifmanLafon06,BerryHarlim16,BerrySauer16b}, including methods developed explicitly for dynamical systems \cite{DellnitzJunge99,Mezic05,RowleyEtAl09,GiannakisMajda12a,BerryEtAl13,GiannakisEtAl15,WilliamsEtAl15,Giannakis17,DasGiannakis17}, decompose signals through pairs of temporal and spatial patterns determined from eigenfunctions of linear operators operating on spaces of scalar-valued functions. This results in low-rank decompositions, with a separable dependence on the spatial and temporal degrees of freedom, and limited efficiency in representing spatiotemporal signals with intermittency in both space and time. Moreover, in the presence of nontrivial spatial symmetries, the patterns recovered by many conventional methods, including PCA, are pure symmetry modes (e.g., Fourier modes in a periodic domain with translation invariance), with minimal dynamical significance and physical interpretability \cite{AubryEtAl93}.     

An alternative approach \cite{GiannakisEtAl17b} is based on the observation that time-evolving spatial patterns generated by dynamical systems can naturally viewed as vector-valued observables on the state space of the system, taking values in a vector space of scalar fields on the spatial domain of interest. Linear operators acting on those spaces, constructed by combining operator-valued kernels for multitask learning \cite{MicchelliPontil05,CaponnettoEtAl08,CarmeliEtAl10} with delay-coordinate maps \cite{PackardEtAl80,Takens81,SauerEtAl91,Robinson05} and Koopman operators \cite{BudisicEtAl12,EisnerEtAl15} of dynamical systems, take advantage of this structure, and lead, through their eigenfunctions, to efficient and physically meaningful decomposition of complex spatiotemporal data. Here, we describe this approach, and demonstrate its efficacy in applications to spatiotemporal data from an oscillator model and sea surface temperature (SST) in the tropical Pacific sector of a comprehensive climate model.

\section{\label{secVSA}Vector-Valued Spectral Analysis}

Consider a continuous-time dynamical system on a manifold $ X $, whose flow map $ \Phi^t : X \mapsto X $, $ t \in \mathbb{ R } $,  possesses a compact invariant set $ A \subseteq X $ and an ergodic invariant probability measure $ \mu $ supported on $ A $. For instance, $ X $ could be the state space of a finite-dimensional ordinary differential equation model, such as the Lorenz 63 system \cite{Lorenz63}, or an inertial manifold of an infinite-dimensional partial differential equation (PDE) model \cite{ConstantinEtAl89,Zelik14}. Let also $ Y $ be a spatial domain (which we assume to have the structure of a compact metric space), equipped with a finite measure $ \nu $, and let $ H_Y = \left \{ f : Y \mapsto \mathbb{ C }, \; \int_Y \lvert f \rvert^2 \, d\nu < \infty \right \} $ be the Hilbert space of square-integrable complex-valued functions on $ Y $ associated with $ \nu $. For our purposes, $ H_Y $ will be the space of physically admissible spatial patterns generated by the dynamical system. In particular, we are interested in a scenario where there is a continuous observation function $  F : X \mapsto H_Y $ mapping each dynamical state $ x \in X $ to a spatial pattern $ F( x ) \in H_Y $. Note that in PDE systems it is oftentimes the case that $ X $ is actually a subset of $ H_Y $; that is, states of the system are scalar fields on $ Y $, and $F$ reduces to an inclusion map. 

In general, we view $ F $ as a \emph{vector-valued observable} of the dynamical system as it takes values in a vector space of functions  $H_Y $. We also require that $ F $ obeys natural integrability conditions; in particular, that it lies in the Hilbert space $ \mathcal{ H } = \left \{ f : X \mapsto H_Y, \; \int_X \lVert f(x) \rVert_{H_Y}^2 \, d\mu(x) < \infty \right \} $ of vector-valued observables which are square-integrable with respect to the invariant measure of the dynamics. Note that $ \mathcal{ H  } $ is isomorphic to the tensor-product space $ H_X \otimes H_Y $, where $ H_X = \left \{ f : X \mapsto \mathbb{ C }, \; \int_X \lvert f \rvert^2 \, d\mu < \infty \right \} $ is the Hilbert space of complex-valued functions on $ X $, square-integrable with respect to $ \mu $. That is, every $ f \in \mathcal{ H }$ admits a decomposition of the form $ f = \sum_{k=0}^\infty u_kv_k $ with $ u_k \in H_Y $ and $ v_k \in H_X $, but, of course, not every element of $ \mathcal{ H } $ is of the tensor-product form $ u v $ for some $ u \in H_Y $ and $v \in H_X$.  

From our perspective, a time-ordered sequence $ h_0, h_1, \ldots, h_{N-1} $ of measurements $ h_n  \in H_Y $, taken at times $t_n = n \tau $ with a sampling interval $ \tau > 0 $, corresponds to samples $ h_n = F( x_n ) $ of the vector-valued observable $ F \in \mathcal{ H } $, taken at the states $ x_n = \Phi^{t_n}( x_0 ) $ of a dynamical trajectory starting at a point $ x_0 \in X $. Given the sequence $ h_n $, and without assuming prior knowledge of the dynamics and/or the structure of the state space or the observation map, our objective is to produce a decomposition 
\begin{equation}
  \label{eqFDecomp}F = \mbox{$\sum_{k=0}^\infty c_k \phi_k$}, \quad c_k \in \mathbb{ R }, \quad \phi_k \in \mathcal{ H },  
\end{equation}    
of the observation map into vector-valued modes $ \phi_k \in \mathcal{ H } $, each capturing a physically meaningful spatiotemporal pattern. 

Before proceeding,  it is worthwhile noting that many data analysis techniques for spatiotemporal data analysis perform such decompositions by effectively imposing a separable structure of the modes $ \phi_k( x, y ) $ on the dynamical/temporal ($x$) and spatial ($y$) degrees of freedom. For instance (ignoring convergence issues due to a finite number of samples $N$), in PCA one computes a set of orthonormal spatial patterns $ u_k \in H_Y $ through the eigenvectors of the spatial covariance operator $ C_Y : H_Y \mapsto H_Y $, and then performs a decomposition of the input signal $ F = \sum_{k=0}^\infty u_k \sigma_k v_k $, where the $ v_k $ are orthonormal temporal patterns in $ H_X $, computed from the projections of the observation map into the spatial modes (viz., $ v_k( x ) = \langle u_k, F( x ) \rangle_{H_Y} $), and the $ \sigma_k $ are equal to the square roots of the eigenvalues of $ C_Y $. Equivalently, the $ u_k $ can be computed from the eigenvectors of a temporal covariance operator acting on functions in $ H_X $. The PCA decomposition is of the class~\eqref{eqFDecomp} with $ c_k = \sigma_k $ and $\phi_k = u_k v_k$, and it follows from the latter relation that the recovered spatiotemporal patterns $ \phi_k $ have a highly restrictive, separable tensor-product structure. Other techniques, including the Nonlinear Laplacian Spectral Analysis (NLSA) algorithm \cite{GiannakisMajda12a} discussed below, lead to significant improvements to the physical interpretability of the recovered modes by incorporating tools such as delay-coordinate embeddings \cite{PackardEtAl80,Takens81,SauerEtAl91,Robinson05} and kernels for manifold learning \cite{ScholkopfEtAl98,BelkinNiyogi03,CoifmanEtAl05,HeinEtAl05,CoifmanLafon06,BerryHarlim16,BerrySauer16b}, while also providing rigorous spectral convergence guarantees on non-smooth invariant sets \cite{DasGiannakis17}. Yet, despite these improvements, these methods also impose a separable or near-separable structure in the recovered spatiotemporal patterns.


Arguably, the shortcomings described above can be attributed to the fact that the spatial and temporal degrees of freedom are treated separately through operators acting on scalar function spaces such as $ H_X $ and $H_Y$. In contrast, our approach is based on operators that act on the natural space $\mathcal{H}$ of vector-valued observables from the outset, without imposing restrictive separability constraints. To construct this class of operators, we take advantage of the fact that, in addition to being isomorphic to $H_X \otimes H_Y$, $\mathcal{H}$ is also isomorphic to the Hilbert space $H_\Omega = \{ f : \Omega \mapsto \mathbb{C}, \; \int_\Omega \lvert f \rvert^2 \, d\rho < \infty \} $ of scalar-valued functions on the product space $ \Omega = X \times Y $, square-integrable with respect to the product measure $ \rho = \mu \otimes \nu $. This means that we can first construct an appropriate integral operator on $ H_\Omega $ based on a scalar-valued kernel on $ \Omega \times \Omega $, and then map that operator to an equivalent operator on $ \mathcal{ H } $, which will be associated with an induced operator-valued kernel \cite{MicchelliPontil05,CaponnettoEtAl08,CarmeliEtAl10}. 

Following ideas introduced in NLSA, we consider a class of kernels $ k_Q : \Omega \times \Omega \mapsto [ 0, 1 ] $, which assign the following measure of similarity to points $ ( \omega, \omega') \in \Omega \times \Omega $, with $ \omega = (x,y) $ and $ \omega' = ( x', y' ) $:
\begin{displaymath}
  k_Q( \omega, \omega' ) = h(d_Q(\omega,\omega')), \quad d_Q^2(\omega,\omega') = \mbox{$\sum_{q=0}^{Q-1}\lvert F((\Phi^{-q\tau}(x),y)) - F((\Phi^{-q\tau}(x'),y')) \rvert^2 / Q$}.  
\end{displaymath}
Here, $ h : \mathbb{R} \mapsto \mathbb{R} $ is a bounded, continuous kernel shape function, and $ Q \geq 1 $ an integer parameter (the number of delays). We nominally utilize a Gaussian shape function, $ h( s ) = e^{-s^2/\epsilon} $, parameterized by a bandwidth parameter $ \epsilon >0 $ tuned automatically as described in \cite{BerryEtAl15}. In effect, $k_Q(\omega,\omega')$ is based on a mean square distance between $Q$-element temporal sequences of the values of the observation map $ F $ at the spatial points $ y,  y' \in  Y $, initialized at the dynamical states $ x, x' \in X $, respectively. Note that $ k_Q $ is manifestly non-separable as a product of kernels on $ X \times X $ and $ Y \times Y $.

Associated with $ k_Q $ is a kernel integral operator $ K_Q : H_\Omega \mapsto H_\Omega $ that maps $ f \in H_\Omega $ to $ K_Qf = \int_\Omega k_Q(\cdot, \omega) f( \omega )\, d\rho(\omega) $. Equivalently, there is a kernel integral operator $\mathcal{K}_Q:\mathcal{H} \mapsto \mathcal{H}$ that acts on vector-valued observables in $\mathcal{H}$ through an operator-valued kernel $l_Q:X\times X \mapsto \mathcal{L}(H_Y)$ that maps every pair $(x,x')$ of dynamical states in $X$ to a bounded integral operator $B = l_Q(x,x')$ acting on functions in $H_Y$ through the formula $ Bh = \int_Y k_Q( x, \cdot, x', y' ) h( y' ) \, d\nu(y') $, $ h \in H_Y $. This leads to the action of $\mathcal{K}_Q$ on a vector-valued observable $f\in\mathcal{H}$ given by $ \mathcal{K}_Q f = \int_X l_Q(\cdot,x) f( x ) \, d\rho(x) $. Next, we normalize $ K_Q $ to a Markov operator $P_Q : H_\Omega \mapsto H_\Omega$ by applying the kernel normalization procedure introduced in the diffusion maps algorithm \cite{CoifmanLafon06}. This induces a corresponding normalized operator $\mathcal{P}_Q : \mathcal{H}\mapsto\mathcal{H} $ acting on vector-valued observables, which is used to identify vector-valued patterns through its eigenfunctions $ \phi_k \in \mathcal{H} $. The latter are used in decompositions of the input signal as in~\eqref{eqFDecomp} with the expansion coefficients $c_k = \langle \phi'_k, F \rangle_\mathcal{H} $ determined through the duals to $ \phi_k $ satisfying $\langle \phi'_j,\phi_k\rangle_{\mathcal{H}}= \delta_{jk}$. A key aspect of this decomposition is that the individual terms $\phi_k$ are not constrained to be products of functions on $ H_X$ and $H_Y$; in fact, when dealing with complex spatiotemporal patterns, the $\phi_k$ will in general be non-separable. 


In \cite{GiannakisEtAl17b}, the following results where obtained: (1) If the system has nontrivial symmetries associated with the action of a group on $ Y$, then the $ \phi_k $ (viewed as scalar-valued functions on $\Omega$, which is possible by the isomorphism $\mathcal{H} \simeq H_\Omega$) are constant on the orbits of an associated group action on $ \Omega $ commuting with the dynamical flow map; as a result, our method can be interpreted as factoring out the natural symmetries (i.e., the symmetries commuting with the dynamics), in the data. (2) When $ Q $ is small, the $ \phi_k $ are approximately constant on the level sets of the observation map $ F $ (viewed as a scalar function on $\Omega$); this is useful for applications requiring estimation of level sets of functions from noisy data, such as topological data analysis \cite{Carlsson09}. (3) In the limit $Q\to\infty$, $\mathcal{P}_\infty$ commutes with the Koopman operator  of the system governing the evolution of vector-valued observables in $\mathcal{H}$; as a result, these operators have common finite-dimensional eigenspaces (by compactness of $\mathcal{P}_\infty$) spanned by observables with a coherent spatiotemporal evolution. (4) Data-driven approximations of these patterns obtained from the time-ordered snapshots $ h_0, \ldots, h_{N-1} $ converge in an appropriate limit of large data, $ N \to \infty $, even if the invariant set $A$ is non-smooth.


\section{\label{secApplications}Applications}

\begin{figure*}
  \includegraphics[width=\linewidth]{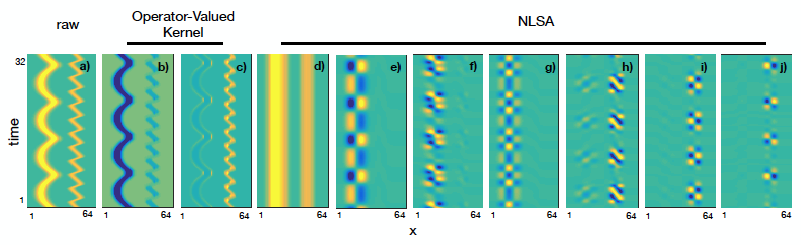}
  \caption{\label{figOscillator} Raw data (a), and spatiotemporal patterns obtained via the vector-valued eigenfunctions of the operator $ \mathcal{P}_Q $ (b, c) and via NLSA (d--j) from the quasiperiodic oscillator model.}
\end{figure*}

Our first application is based on data generated by an ergodic dynamical system whose state space is the 2-torus,  $ X = \mathbb{T}^2$. On $ X $, the dynamics is governed by a variable-speed quasiperiodic rotation, $ \Phi^t( (\theta_1, \theta_2)) = (\theta_1+\alpha_1(t), \theta_2+ \alpha_2(t) ) $, where $ \alpha_1, \alpha_2 $ are periodic functions such that the ratio $ r = T_1 / T_2 $ of the corresponding periods is irrational. This system was also employed in \cite{Giannakis17}, where it is described in more detail. Here, we set $ r = \sqrt{7} $, and consider the nonlinear observation function $ F( ( \theta_1, \theta_2 ) ) = g_1 + g_2 \in H_Y $,  where $ Y = \mathbb{ T}^1 $ is a one-dimensional periodic domain, and $ g_1( y ) = e^{- (y - \cos(\theta_1) /11 - 0.3)^2/0.005} $, $ g_2( y ) = e^{- (y- \cos(\theta_2))/16 - 0.7)^2/0.002 } / r $. The resulting spatiotemporal signal, displayed in Figure~\ref{figOscillator}(a), features two spatially localized periodic disturbances with a non-separable structure in time and space and rationally-independent periods. As shown in Figure~\ref{figOscillator}(b, c) the vector-valued eigenfunctions of  the operator $ \mathcal{P}_Q $ determined from this dataset (using $ Q = 32 $ delays at a timestep $ \tau \approx 0.05 \, T_2 $) individually capture the two components of the oscillatory spatiotemporal signal. In contrast, the NLSA patterns in Figure~\ref{figOscillator}(d--j) exhibit a low-rank, separable behavior in space and time, and while they split into two families, (e--h) and (h--j), associated with the two oscillatory disturbances, they are not representative of the structure of the signal in space.    

\begin{figure*}
  \centering \includegraphics[width=.6\linewidth]{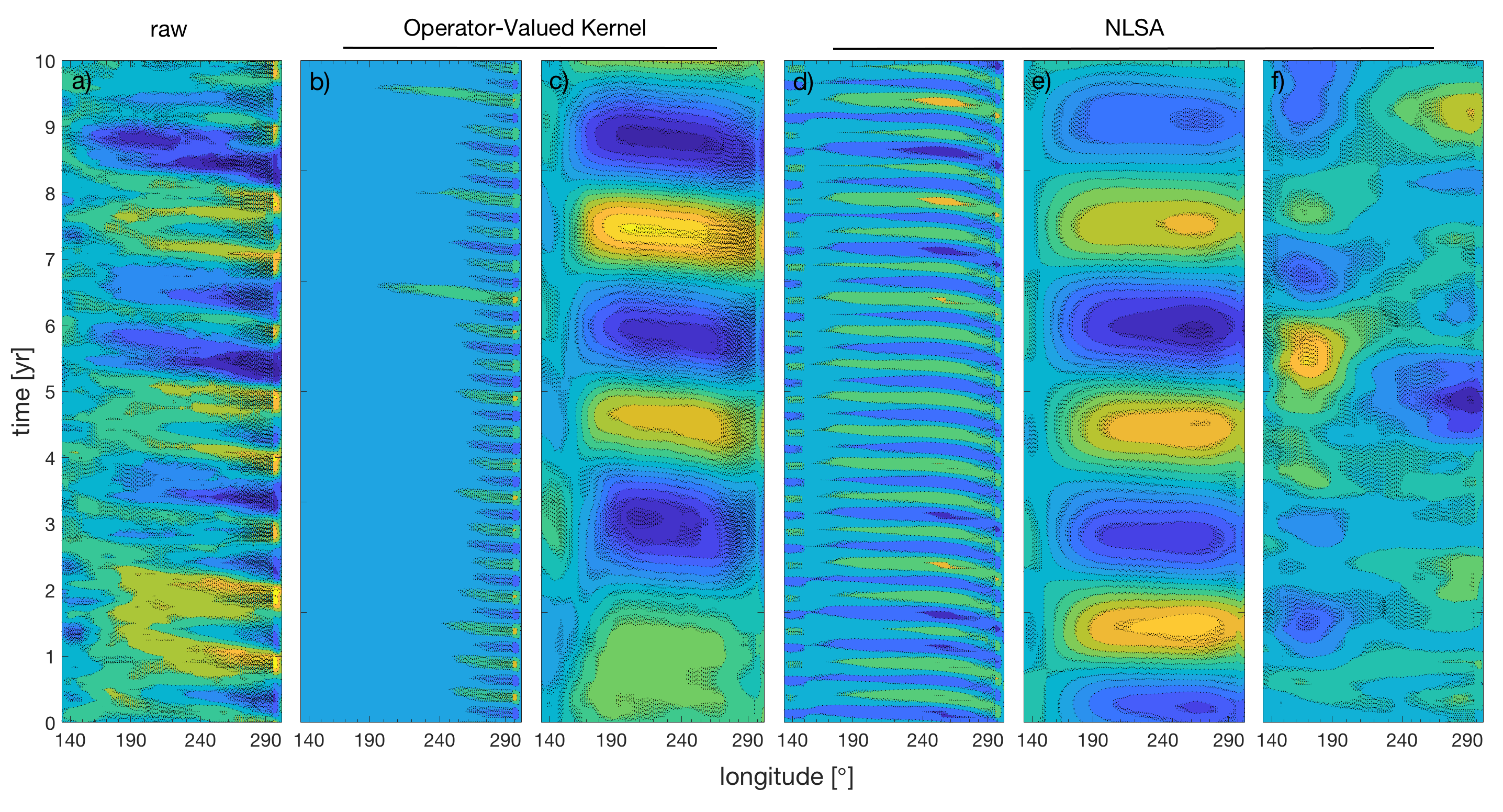}
  \caption{\label{figENSO}As in Figure~\ref{figOscillator}, but for patterns extracted from equatorial Pacific SST data from CCSM4.}
\end{figure*}

Next, we examine SST data generated by the CCSM4 climate model \cite{GentEtAl11}, sampled in the Pacific Ocean (spatial domain $ Y = \{ \theta \in [ 140^\circ\text{E}, 65^\circ\text{W}]\} $, where $ \theta$ denotes longitude) on the equator at a timestep $ \tau = 1 $ month. As shown in Figure~\ref{figENSO}(a), this domain exhibits SST variability on seasonal to multidecadal timescales (\cite{SlawinskaGiannakis17,GiannakisSlawinska17}, and references therein), including the El Ni\~no Southern Oscillation (ENSO)---the dominant mode of interannual ($\sim \text{3--5}$ year) variability of the coupled atmosphere-ocean system with considerable global influences on weather and climate patterns. ENSO is captured here through distinct families of modes via both vector-valued eigenfunctions of $ \mathcal{ P }_Q $ (Figure~\ref{figENSO}(c)) and NLSA (Figure~\ref{figENSO}(e, f)) with $ Q = 48 $ (4 years), but there are important differences between the two methods. In particular, the vector-valued eigenfunctions provide a more realistic representation of ENSO diversity \cite{LeeMcPhaden10}; that is, the tendency of ENSO to exhibit variations from event to event in an aperiodic manner. The vector-valued approach requires only one pattern (Figure~\ref{figENSO}(c)) to represent eastern and central Pacific ENSO events (e.g., years 8--10 and 2--4, respectively). On the other hand, NLSA requires several ENSO modes, including the patterns in Figure~\ref{figENSO}(e, f), to reproduce ENSO diversity. Other modes (Figure~\ref{figENSO}(b, d)) exhibit semiannual variability, which is pronounced in the tropics. The semiannual pattern from the vector-valued approach in Figure~\ref{figENSO}(b) is localized in eastern Pacific longitudes associated with the South American monsoon, whereas the corresponding pattern from NLSA  in Figure~\ref{figENSO}(d)  is more uniformly distributed over the domain.
  

\subsubsection*{Acknowledgments}

D.~G.\ acknowledges support from NSF EAGER grant 1551489, ONR YIP grant N00014-16-1-2649, NSF grant DMS-1521775, and DARPA grant HR0011-16-C-0116. J. S.\ and A.~O.\ acknowledge support from NSF EAGER grant 1551489. Z.~Z.\ received support from NSF grant DMS-1521775.

\bibliography{bibliography}

\end{document}